\documentclass[12pt,english]{article}
\pdfoutput=1
\usepackage[T1]{fontenc}
\usepackage[latin1]{inputenc}
\usepackage{geometry}
\usepackage{rotating}
\usepackage{graphicx}
\usepackage{setspace}
\makeatletter

\providecommand{\tabularnewline}{\\}

\usepackage{graphicx}

\usepackage[round, comma, sort&compress, authoryear]{natbib}

\usepackage{xspace}

\newcommand{\pdb}{\textsc{pdb}\xspace}

\newcommand{\gabb}{\textsc{gabb}\xspace}
\newcommand{\rmsd}{\textsc{rmsd}\xspace}

\newcommand{\rapper}{\textsc{rapper}\xspace}
\newcommand{\rappertk}{\textit{Rapper}tk\xspace}
\newcommand{\mdsa}{\textsc{mdsa}\xspace}
\newcommand{\cns}{\textsc{cns}\xspace}

\newcommand{\Ang}[1]{${#1}$\AA\xspace}

\date{}

\usepackage{babel}
\makeatother
\begin{document}

\title{Crystallographic modelling of protein loops and their heterogeneity
with \rappertk\footnote{This document is very similiar to a chapter in SG's PhD thesis submitted in Sept.2007 to the University of Cambridge, England.}}

\author{Swanand Gore and Tom Blundell\\
\{swanand,tom\}@cryst.bioc.cam.ac.uk\\
Department of Biochemistry, University of Cambridge\\
Cambridge CB2 1GA England}

\maketitle
\begin{abstract}
\textbf{Background} All-atom crystallographic refinement of proteins
is a laborious manually driven procedure, as a result of which, alternative
and multiconformer interpretations are not routinely investigated. 

\textbf{Results} We describe efficient loop sampling procedures in
\rappertk~ and demonstrate that single loops in proteins can be
automatically and accurately modelled with few positional restraints.
Loops constructed with a composite \cns/\rappertk~ protocol consistently
have better $R_{free}$ than those with \cns~ alone. This approach
is extended to a more realistic scenario where there are often large
positional uncertainties in loops along with small imperfections in
the secondary structural framework. Both ensemble and collection methods
are used to estimate the structural heterogeneity of loop regions.

\textbf{Conclusion} Apart from benchmarking \rappertk~ for the all-atom
protein refinement task, this work also demonstrates its utility in
both aspects of loop modelling - building a single conformer and estimating
structural heterogeneity the loops can exhibit.
\end{abstract}

\newpage
\section{Introduction}

X-ray crystallography has been the most popular protein structure
determination technique of both pre- and post-genomic eras. The challenges
of macromolecular crystallography are manifold - after the difficult
steps of expression, purification, crystallization and data collection,
there remains the final and important task of data interpretation
in order to build a model which explains the observed diffractions.
Structural interpretation requires overcoming the phase problem and
often starts with partial and incorrect phases. Typically, semi-automatic
iterative refinement is carried out, gradually improving the model's
quality as indicated by the $R$ and $R_{free}$ factors as well as
decrease in covalent geometry and excluded volume violations. Although
excellent softwares like CCP4 (\cite{ccp4}), Phenix (\cite{phenix})
and {\small \cns} (\cite{cns}) make this task possible, the structure
refinement procedure remains manually-driven hence laborious and subjective.
Due to this, the heterogeneity in structural interpretation of diffraction
data is often ignored in favour of a single-conformer isotropic B-factor
model.

Protein structure is important for its function. But very stable,
rigid proteins cannot exhibit enzymatic activity. This suggests that
proteins have to be stable enough to retain their fold yet dynamic
enough to be functional. Both experimental and computational studies
indicate that single-conformer interpretation of crystallographic
data is not adequate to capture the native state dynamics which is
largely conserved even in a crystal owing to its high solvent content
(\cite{xrayLigReverseBinding}, \cite{crystalfreedom}). Reporting
a multiconformer interpretation of data will make use of the structure
less misleading, especially in the analyses that depend on geometry
such as shapes of binding sites, orientations of sidechains, detection
of non-covalent interactions and so on. While multiple interpretations
are necessary, they should be free from any bias such as that introduced
when different crystallographers solve the same diffraction data.
Multiconformer interpretation will be greatly facilitated by automated
methods.

Thus multiple persuasive justifications emerge for automating the
protein crystallographic refinement task: (a) capturing the dynamics
of protein in the crystalline state (b) removing subjective bias from
the refinement process and (c) reducing the need for precious human
resource. But this goal is hard to achieve in practice. The under-determined
nature of the problem (number of independent observation < number
of parameters) prevents a straightforward solution by minimization.
Even when sufficient restraints exist, minimization methods like conjugate
gradient, steepest descent etc. suffer from the problem of local minima.
Hence use of well-known features of proteins is unavoidable. Automatic
pattern recognition in electron density is very successful in presence
of high resolution data and good phases because it looks for such
features (\cite{arpwarp}). But at medium resolution or given poor
phases, this strategy can get misled.

Our recent efforts with automated crystallographic refinement started
with {\small \rapper}, which is a conformation sampling program for
proteins and uses a genetic algorithm cum branch-and-bound ({\small \gabb})
algorithm. \cite{xrayHetInaccu} showed that multiple interpretations
similar to the deposited structure are possible given the deposited
data, and the divergence in interpretation is correlated to resolution.
With {\small \rapper}, it was demonstrated (\cite{rapperKnowledgeXray})
that when a protein structure is approximately known, it can be refined
to native-like quality, unlike {\small \mdsa} in {\small \cns} which
may get stuck in local minima. Fundamental features of {\small \rapper}
responsible for avoidance of local minima traps were (a) use of fine-grained,
propensity-weighted $\phi-\psi$ maps for backbone sampling (b) use
of backbone-dependent rotameric libraries (c) use of ideal Engh and
Huber covalent geometry (d) mild use of electron density and positional
restraints to guide the sampling process. Later \cite{rapperLowResolution}
demonstrated that a low-resolution dataset can be rescued and interpreted
semi-automatically to obtain structure of a system with great biological
significance.

\cite{rapperKnowledgeXray} observed that automatic refinement becomes
less satisfactory as positional restraints become weaker: the structures
could not be refined if the initial $C_{\alpha}$ perturbation was
of order of \Ang{3} or more. This is not unexpected because larger
positional restraints dilute the information and would make the search
harder. But often a practical problem encountered in crystallography
is that of missing loops, i.e. knowing loop regions with far less
$C_{\alpha}$ positional certainty than the regions with regular secondary
structure. By definition, loops exhibit rich variability in backbone
torsion angles. They are thought to be more dynamic than the protein
secondary structural framework and also functionally more interesting.
Thus it is important to use the available restraints as efficiently
as possible to build loops despite weaker electron density and greater
positional uncertainty while tolerating small positional errors in
the framework.

After determining a single-conformer loop structure, the second important
challenge is to estimate the structural variability of the loops.
It is easy to see by generating artificial data that existence of
structural heterogeneity for a loop results in confusing electron
density. In general, partial occupancies result in weaker density
than full occupancy. Sidechains of the same residue may occupy different
density contours. Overlaps in conformations may lead to significant
loss of shape information. These challenges can be expected to make
the task only harder for minimization-based programs when refining
a multiconformer structure.

Following the reformulation of {\small \rapper} as a versatile modular
software called \rappertk (\cite{rappertk}), it was essential to
benchmark its performance for all-atom protein crystallographic refinement.
Hence the first result in this work is the all-atom knowledge-based
crystallographic refinement given the positional restraints for the
entire protein, establishing that a similar result as {\small \rapper}
(\cite{rapperKnowledgeXray}) can be achieved. We then demonstrate
that single loops in proteins can be reconstructed to a high quality
with \rappertk using little positional information. This case is
then extended to include all loop regions and a small error in the
framework to show that the composite {\small \cns}/\rappertk refinement
approach is suitable in a realistic scenario. Finally, we ask whether
single-loop heterogeneity can be modelled with collections of independently
generated models or ensembles of conformers.

\section{Methods}

\subsection{Overview of iterative refinement}

Each step in refinement procedure consists of: (a) identification
of residues which do not fit well into density, (b) finding contiguous
bands of such residues, (c) rebuilding the bands with knowledge-based
conformational sampling within restraints, (d) optimal sidechain placement
of rebuilt sidechains and (e) refining the resulting model with {\small \cns}.

The fit of a set of atoms to electron density is calculated as the
correlation coefficient between the $\sigma_{A}$-weighted omit map
and $F_{c}$ map for a region around \Ang{1} of the atoms. The maps
used are both generated by {\small \cns} refinement script, hence
they are described on the same grid. Following \cite{RrealJones1991}
and \cite{rapperKnowledgeXray}, the correlation coefficient between
the maps is calculated on the grid neighbourhood around atoms of interest:

\begin{equation}
CorrCoef=\frac{\sum\sigma_{omit}\sigma_{c}}{\sqrt{\sum\sigma_{omit}^{2}\sum\sigma_{c}^{2}}}\label{eqChisqCorrCoeff}\end{equation}

If the correlation is below $0.9$, the atoms are flagged for rebuilding.
Correlation is calculated on all atoms in residues and then on mainchains
only, sidechains only and peptide atoms only. Ill-fitting sidechains
are marked for sidechain reassignment whereas residues with ill-fitting
peptide or mainchain or all-atoms are marked for all-atom reconstruction.

Once the residues are flagged for all-atom rebuilding, contiguous
bands are identified and marked either as N-terminal, C-terminal or
intermediate. Bands are then sampled in random order using the PopulationSearch
algorithm. Each band is attempted 5 times and left as it was if it
cannot be sampled within given restraints. Previously sampled bands
are considered while sampling later bands. N and C terminal bands
are built using forward and reverse techniques and weighted sampling
of $\phi-\psi$ propensities. Building of intermediate regions is
described in a later section.

Once all bands are sampled, all the resampled sidechains are reassigned
using the optimal sidechain placement procedure described elsewhere
(\cite{opsax}).

\subsection{Electron Density Ranker}

Generally a single conformer model would refine better if its occupation
of the $2F_{o}-F_{c}$ map is better - so a model within $1\sigma$
contour is more reliable than $0.5\sigma$. Thus, on output of each
builder, a binary electron density restraint can be applied with a
$\sigma$-level cutoff. But the quality of map is not uniform over
all residues and hence such binary restraint is useful only for ensuring
that model remains within positive electron density. Hence, in addition
to that restraint, an analog ranker is used for electron density that
ranks the possibilities and chooses the better ones. In the population
search algorithm, the ranker asks more children to be generated at
each conformation extension step than the population size (typically
5 times more), ranks them and chooses top-ranking ones to fill the
conformation pool. The ratio of number of children generated to the
population size is termed as the enrichment ratio. The electron density
ranking scheme calculates score of a set of atoms by summing up the
$\sigma$ values in a \Ang{1} region around their coordinates. The
effective $\sigma$ value is calculated by penalising the negative
$\sigma$ and flattening the peaks by upper cutoff, the latter for
better recognition of shape of density rather than spikes, say due
to waters or ions. In addition to filtering children at each step,
the ranker also chooses the best member from the conformational pool
generated, which is returned as the sampled model for the band.

\subsection{Symmetry-related clash cheking}

As described in \cite{rappertk}, \rappertk uses geometric caching
implemented as Clashchecking grid for efficiently deciding whether
atomic van der Waals spheres are overlapping. This excluded-volume
restraint rules out many unproductive sampling trajectories. When
loop positions are largely uncertain, the existence of symmetry-related
images of the molecule around it acts as an excluded volume restraint
to loop sampling. \rappertk uses the Clipper (\cite{clipper}) libraries
for crystallographic computing for symmetry-related calculation. Clashchecking
grid uses Clipper's Spacegroup class and symmetry operators therein
to calculate the images of atoms to be added into the grid. Images
within \Ang{20} of the bounding box of given protein coordinates
are considered. First the grid looks for any clashes between sampled
coordinates and their images. Then it is verified that they do not
clash against the rest of the coordinates or their images. In case
of no clashes, the new coordinates and their images are added to the
grid. Removing coordinates from the grid removes their images too.

\subsection{Loop closure}

The typical incremental sampling step in \rappertk builds $C_{\alpha}^{i}$
and $(i-1)^{th}$ sidechain in the forward mode or $C_{\alpha}^{i}$
and $(i+1)^{th}$ sidechain in the reverse mode. In this context,
loop closure can be formulated as finding the locations of mainchain
atoms $\{ C^{i-1}$, $O^{i-1}$, $N^{i}$, $C_{\alpha}^{i}$, $C^{i}$,
$O^{i}$, $N^{i+1}\}$ and sidechains of $(i-1)^{th}$,$i^{th}$ and
$(i+1)^{th}$ residues. Seamless loop closure of this kind is challenging
because many conditions need to be met: (a) the covalent angles and
lengths should be correct (b) $\phi,\psi$ states of 3 residues should
be in the allowed regions (c) two $\omega$ angles should be adopt
cis or trans conformation, but not be restricted to one or the other
(d) 3 sidechains should be rotameric and (e) van der Waals restraints
should be obeyed.

A sampling procedure was devised for meeting conditions (a), (b),
(c) and (d), while (e) is met using clash-checking restraints. The
sampling procedure is similar to the one described in \cite{rappertk},
but modified to meet the cis $\omega$ state too. First, the two $\omega$
angles are sampled, leading to the center, plane and radius of the
circle on which the middle $C_{\alpha}$ is sampled. The circle is
uniformly sampled. For each sample, the $\{ r,\alpha,\theta,\}-\{\phi,\psi,\omega\}$
mapping is used to build the mainchain atoms. Then the three sidechains
are sampled from a rotamer library. Sampling is continued until a
conformation satisfying all restraints is found.

The problem with this sampling is that the restraint density abruptly
increases at loop closure because it is not clear how to back-propagate
the geometric requirements (a) and (c). Due to this, the sampling
procedure fails often and takes a long time to find a valid sample.
Often an incorrect conformation is built in case of imperfect density
because sampling of $\phi,\psi,\omega$ is not propensity-weighted.
Hence after significant experimentation with this approach, it was
abandoned in favour of a simpler approximate approach.

In the simpler approach, the loop closure is formulated as finding
the coordinates of mainchain atoms $\{ C^{i}$, $O^{i}$, $N^{i+1}\}$
and sidechains of residues $i$ and $i+1$. A $\phi$ sampler is used
to build the $C^{i}$ atom which is required to lie between $0.5$
to \Ang{2} from the $N^{i+1}$ atom. Covalent angles $N^{i}-C_{\alpha}^{i}-C^{i}$
and $C^{i}-N^{i+1}-C_{\alpha}^{i+1}$ are restrained to lie between
$90^{o}$ and $150^{o}$. $\omega$ dihedral angle $C_{\alpha}^{i}-C^{i}-N^{i+1}-C_{\alpha}^{i+1}$
is allowed a maximum deviation of $30^{o}$ from cis or trans conformation.
Two sidechains are sampled for each $C^{i}$ sampled. It is observed
that it is more efficient to close a loop with this method than the
previous.

\subsection{Both-sided sampling}

Bands to be rebuilt can be of three types: the N terminal band, the
C terminal or intermediate. For the C and N terminal bands, forward
and reverse sampling are used respectively. For the intermediate regions,
the most efficient way is to use a both-sided sampling approach as
opposed to only forward sampling. As explained previously (see \cite{rappertk}
in the context of $\beta$-sheet sampling), in both-sided sampling,
residues are sampled in the order $i,k,i+1,k-1,i+2,k-2,...$. In case
of forward or reverse sampling, only a weak loop-closure distance
restraint informs the sampling process of the other end of the loop,
but with both-sided sampling, information at both N and C termini
is actively used. A distance restraint is used between $C_{\alpha}$
atoms at the same sequence distance from both termini, so that the
chance of loop closure remains high despite both sided sampling. Initial
experiments with crystallographic loop building clearly showed that
refinement was better with both-sided sampling than forward-only sampling,
especially with larger loops and weaker positional restraints. Thus,
in this work, we have used forward-only, backward-only and both-sided
sampling for C terminal, N-terminal and intermediate bands respectively.

\subsection{Multiconformer sampling}

This type of sampling constructs many conformations of the same band
simultaneously. Instead of incrementally sampling one band, multiple
models of the band are extended simultaneously. This is achieved by
re-implementing the PopulationSearch algorithm in its plural form
in which each builder is replaced by a set of builders that have same
input and output atoms in different models. Clashchecking is not performed
across models. Electron density ranker uses the combined output of
a set of corresponding builders to calculate the score of a child
conformation. Due to this, the possibility of getting attracted into
higher density is reduced and the chance of occupying the density
generated due to genuine heterogeneity increases. The disadvantage
of this kind of sampling is the obvious increase in conformational
freedom and execution time.

\section{Results}

\subsection{Reproducing {\small \rapper}/{\small \cns} refinement}

The utility of knowledge-based refinement has been demonstrated by
\cite{rapperKnowledgeXray} with automatic refinement of perturbed
starting structures of 9ILB and 1KX8 to an $R_{free}$ almost same
as the deposited structure. When that refinement protocol was closely
reproduced in \rappertk, very similar results were obtained. Five
proteins were selected in the \Ang{2}-\Ang{3} resolution range from
the {\small \pdb}: 9ILB (\cite{ref9ILB}), 1KX8 (\cite{ref1KX8}),
1MB1 (\cite{ref1MB1}), 1BYW (\cite{ref1BYW}) and 1RN7 (\cite{ref1RN7}).
Five perturbed structures were generated for each of them within \Ang{2}
$C_{\alpha}$ and \Ang{3} sidechain centroid restraints respectively.
20 rounds of both {\small \cns}-only and {\small \cns}/\rappertk
refinement protocols were performed on these starting models to obtain
the $R_{free}$ statistics summarized in Table \ref{fullprotTable}.
$R_{free}$ figures reported for 9ILB and 1KX8 by \cite{rapperKnowledgeXray}
for same restraints were $0.27(0.01)$ and $0.32(0.01)$ respectively
- the corresponding statistics observed here are comparable.

\begin{table}

\caption{Full-protein testset and refinement statistics for 5 starting models
generated within \Ang{2} $C_{\alpha}$ and \Ang{3} sidechain restraints.}

\begin{center}\begin{sideways}
\begin{tabular}{|c|c|c|c|c|c|c|c|c|}
\hline 
PDB Id&
Resolution&
\#AA&
Baseline$^{a}$&
CNS-only&
\multicolumn{4}{c|}{CNS/\rappertk}\tabularnewline
&
 (\Ang{})&
&
$R_{free}$&
$R_{free}$&
$R_{free}$&
AA-RMSD$^{b}$&
$\chi_{1}$$^{c}$&
$\chi_{12}$$^{d}$\tabularnewline
&
&
&
&
&
&
\multicolumn{3}{c|}{(pdb vs. models, among models)}\tabularnewline
\hline
\hline 
9ILB&
2.28&
153&
0.225&
0.266 (0.016)&
0.253 (0.011)&
1.29 (0.08)&
75.2 (2.63)&
53.0 (4.64)\tabularnewline
&
&
&
&
&
&
0.585&
79.100&
55.100\tabularnewline
\hline 
1KX8&
2.80&
99&
0.294&
0.317 (0.018)&
0.310 (0.010)&
0.60 (0.02)&
76.2 (1.93)&
50.2 (4.06)\tabularnewline
&
&
&
&
&
&
0.551&
77.800&
57.500\tabularnewline
\hline 
1MB1&
2.10&
98&
0.292&
0.333 (0.021)&
0.302 (0.003)&
0.60 (0.02)&
75.8 (2.31)&
56.4 (2.05)\tabularnewline
&
&
&
&
&
&
0.598&
78.300&
59.600\tabularnewline
\hline 
1BYW&
2.60&
110&
0.292&
0.342 (0.020)&
0.316 (0.013)&
0.58 (0.06)&
78.6 (3.32)&
57.4 (5.60)\tabularnewline
&
&
&
&
&
&
0.595&
79.700&
58.400\tabularnewline
\hline 
1RN7&
2.50&
112&
0.270&
0.366 (0.054)&
0.318 (0.009)&
0.61 (0.04)&
81.0 (2.09)&
53.2 (1.93)\tabularnewline
&
&
&
&
&
&
0.604&
80.300&
54.500\tabularnewline
\hline
\hline 
\multicolumn{9}{|c|}{}\tabularnewline
\multicolumn{9}{|c|}{\begin{minipage}[c]{200mm}%
$^{a}$Baselines $R_{free}$ values are computed by running a long
CNS-only refinement starting from the deposited structure.

$^{b}$All-atom RMSDs are in \Ang{} units. First two numbers are
the mean and standard deviations of RMSDs between the PDB and 5 models.
The third number is the mean of RMSDs among 5 models based on $^{5}C_{2}=10$
pairwise comparisons.

$^{c}$$\chi_{1}$ accuracy of a protein structure with respect to
another is defined as the percentage of sidechains in the former that
have a $\chi_{1}$ differing by less than $40^{o}$ from the corresponding
sidechain in the latter. First two numbers are the mean and standard
deviation of the $\chi_{1}$ accuracies between 5 models and the deposited
structure. The third number is the mean of $\chi_{1}$ accuracies
from $^{5}C_{2}=10$ pairwise comparisons among the 5 models.

$^{d}$Same as $^{c}$, but with $\chi_{1}$ and $\chi_{2}$ both
instead of only $\chi_{1}$.\end{minipage}%
}\tabularnewline
\multicolumn{9}{|c|}{}\tabularnewline
\hline
\end{tabular}
\end{sideways}\end{center}

\label{fullprotTable}
\end{table}

Table \ref{fullprotTable} shows the variation in all-atom {\small \rmsd}
and $\chi_{1},\chi_{12}$ values as function of resolution. The reported
{\small \rmsd} is the average of pairwise unsuperposed {\small \rmsd}
between the deposited and each of the composite models and thus can
be said to indicate the inaccuracy in retrieving the deposited model
from an approximate starting model. This inaccuracy does not seem
to be sensitive to the resolution, suggesting that at least in the
\Ang{2.1}-\Ang{2.8} resolution range, an approximate model can be
corrected to a similar quality with respect to the deposited one irrespective
of the resolution.

When the models are compared among themselves in a pairwise manner,
the average {\small \rmsd} and $\chi_{1},\chi_{1,2}$ figures can
be said to represent heterogeneity. This heterogeneity is slightly
lower than the inaccuracy, but the difference is insignificant, i.e.
each model is as far from the deposited structure as from any other
model. Recently the heterogeneity defined similarly has been suggested
to be the minimum uncertainty expected in the coordinates of a single-conformer
model of that structure (\cite{minUncertaintyXrefine}). 

An ideal refinement method should start with approximate models and
yield a set of high heterogeneity models each of which agrees at least
as well with the data as the deposited structure, i.e. a combination
of results in \cite{xrayHetInaccu} and \cite{rapperKnowledgeXray}.
Clearly, the protocol used is similar to \cite{rapperKnowledgeXray}
and perhaps expectedly, does not yield greater heterogeneity than
inaccuracy. But when the models are assigned partial occupancies and
combined to create a multiconformer model (Fig.\ref{LoopCollectionRfree},
Section \ref{collectionRfreeSection}), the collection $R_{free}$
values for 1MB1, 1BYW and 1KX8 drop significantly by $1.5-2\%$ than
the deposited structure. This drop suggests that perhaps structural
heterogeneity is captured to some extent.

\subsection{Rebuilding missing loops}

Five structures of various resolutions and no obvious homology were
selected from the {\small \pdb}: 1MB1 (\cite{ref1MB1}), 1BYW (\cite{ref1BYW}),
1KXB (\cite{ref1KXB}), 1RN7 (\cite{ref1RN7}) and 2DBO (\cite{ref2DBO}).
All structures have a single continuous peptide chain between 100-200
residues and no ligands. For each structure, a loop at least 10 residues
long was chosen for rebuilding (Table \ref{loopdataset}). Unlike
the previous exercise, there are no positional restraints on loop
sidechains. $C_{\alpha}$ atoms are positionally restrained, in the
first case with \Ang{5} restraints and later with \Ang{10} restraints.
The loops were rebuilt using the both-sided loop sampling and loop
closure techniques within the iterative refinement protocol. For 4
of 5 loops considered, the \Ang{5} perturbation of the loop can be
corrected to within one point of the baseline $R_{free}$. In the
\Ang{10} case, this performance drops marginally to two points from
the baseline $R_{free}$ for the same cases.

\begin{figure}

\caption{Loop building exercise for the 1MB1 loop with \Ang{10} $C_{\alpha}$
restraints. Top panel shows the loop in the deposited structure (green)
and starting models generated for it. Middle panel shows the best
$R_{free}$ models (slate) obtained during the CNS-only refinement.
Bottom panel shows the CNS/\rappertk models (magenta) and the loop
in the deposited structure (green) in all-atom representation.}

\begin{center}\includegraphics[%
  width=100mm]{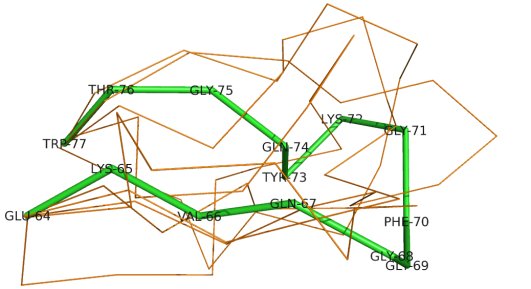}\end{center}

\begin{center}\includegraphics[%
  width=100mm]{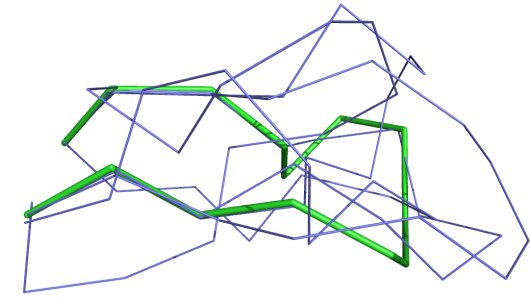}\end{center}

\begin{center}\includegraphics[%
  width=100mm]{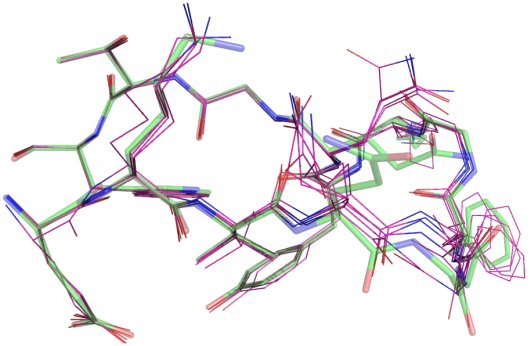}\end{center}

\label{loopfig1MB1}
\end{figure}

\begin{figure}

\caption{Loop building exercise for the 1RN7 loop with \Ang{5} $C_{\alpha}$
restraints. Panels arranged in a similar way to Fig.\ref{loopfig1MB1}.}

\begin{center}\includegraphics[%
  width=100mm]{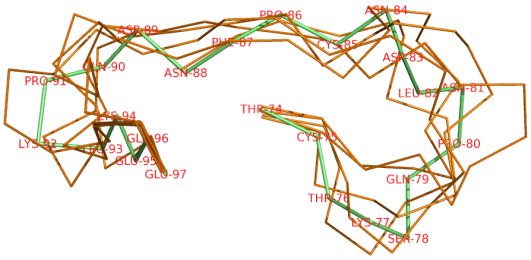}\end{center}

\begin{center}\includegraphics[%
  width=100mm]{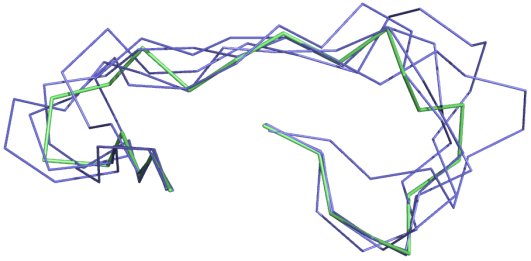}\end{center}

\begin{center}\includegraphics[%
  width=100mm]{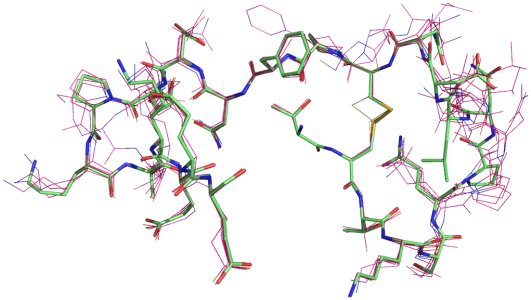}\end{center}

\label{loopfig1RN7}
\end{figure}

Fig.\ref{loopfig1MB1} shows the large difference in the quality of
{\small \cns}-only and {\small \cns}/\rappertk refinement protocol.
Every starting model refines to a structure very similar to the deposited
using the composite protocol whereas it gets trapped in local minima
during {\small \cns}-only refinement. The 1RN7 case (Fig.\ref{loopfig1RN7})
is unsatisfactory due to a difficult 5-residue segment in the loop
(Pro-80, Asn-81, Leu-82, Asp-83, Asn-84). As noted by \cite{ref1RN7},
the density for this segment is confusing, perhaps due to underlying
heterogeneity, and consistently misleads the band sampling into a
conformation different from the deposited.

\begin{table}

\caption{Dataset for loop building and refinement statistics.}

\begin{center}\begin{sideways}
\begin{tabular}{|c|c|c|c|c|c|c|c|c|}
\hline 
PDB Id&
Resolution&
\#AA&
Loop range&
Baseline &
\multicolumn{2}{c|}{$R_{free}$(CNS-only)}&
\multicolumn{2}{c|}{$R_{free}$(CNS/\rappertk)}\tabularnewline
&
 (\Ang{})&
&
and size&
$R_{free}$&
\Ang{5}&
\Ang{10}&
\Ang{5}&
\Ang{10}\tabularnewline
\hline
\hline 
1MB1&
2.1&
98&
66-75 (12)&
0.292&
0.331 (0.024)&
0.368 (0.022)&
0.304 (0.013)&
0.307 (0.005)\tabularnewline
\hline 
1BYW&
2.6&
110&
115-124 (12)&
0.292&
0.329 (0.006)&
0.335 (0.008)&
0.299 (0.012)&
0.314 (0.008)\tabularnewline
\hline 
1KXB&
2.9&
158&
206-218 (15)&
0.283&
0.326 (0.007)&
0.337 (0.010)&
0.290 (0.008)&
0.293 (0.017)\tabularnewline
\hline 
1RN7&
2.5&
112&
76-95 (22)&
0.270&
0.347 (0.018)&
0.366 (0.022)&
0.311 (0.009)&
0.316 (0.013)\tabularnewline
\hline 
2DBO&
2.76&
148&
83-98 (18)&
0.289&
0.332 (0.015)&
0.354 (0.006)&
0.291 (0.010)&
0.291 (0.008)\tabularnewline
\hline
\end{tabular}
\end{sideways}\end{center}

\label{loopdataset}
\end{table}

\subsection{Framework and loops}

\begin{figure}

\caption{Framework/loop exercise for 1BYW. The deposited structure is shown
as thick green ribbon and starting structures as brown ribbon in the
top panel. Middle panel shows the best $R_{free}$ structures from
CNS-only trajectories (slate) and bottom panel shows those from from
the CNS/{\small \rappertk} trajectories (magenta).}

\begin{center}\includegraphics[%
  width=100mm]{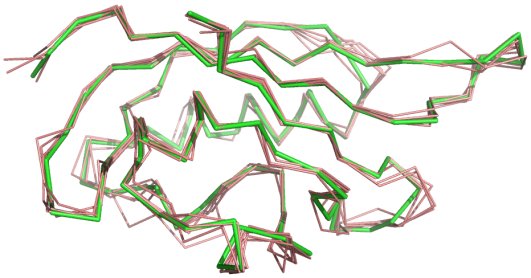}\end{center}

\begin{center}\includegraphics[%
  width=100mm]{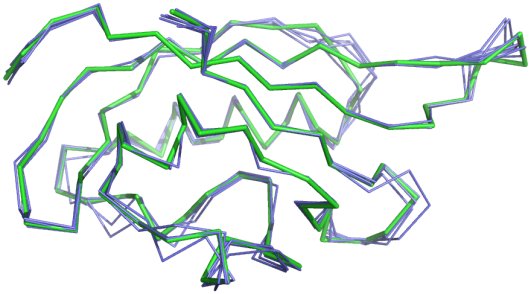}\end{center}

\begin{center}\includegraphics[%
  width=100mm]{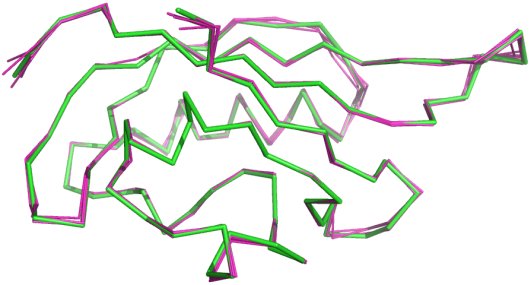}\end{center}

\label{loopfig1BYWfr}
\end{figure}

Perhaps a more frequently encountered scenario than the previous two
is the one in which both the secondary structure and loops have positional
uncertainty. In such cases, the loop regions invariably are more unreliable
than the secondary structure framework. In order to simulate this
scenario, the framework was restrained to \Ang{1} $C_{\alpha}$ and
\Ang{3} sidechain centroid restraints, whereas loops were restrained
to \Ang{3} $C_{\alpha}$ restraints and no sidechain restraints.
Five models were built for each protein and then iteratively refined
using both {\small \cns}-only and {\small \cns}/{\small \rappertk}
protocols. The refinement statistics are summarized in Table \ref{frameloopTable}.
Note that the refinement composite refinement statistics are not as
good as in the previous exercises, but still better than the {\small \cns}-only
refinement. Fig.\ref{loopfig1BYWfr} shows a typical contrast between
the {\small \cns}-only and the composite refinement protocols in
this scenario.

\begin{table}

\caption{Dataset and refinement statistics for the loop-framework refinement}

\begin{center}\begin{tabular}{|c|c|c|c|c|c|c|}
\hline 
PDB&
Resolution&
\#AA&
Loops&
\multicolumn{3}{c|}{$R_{free}$ }\tabularnewline
&
 (\Ang{})&
&
\#AA (\%)&
baseline&
CNS-only&
CNS/{\small \rappertk}\tabularnewline
&
&
&
&
&
Mean (Std.Dev.)&
Mean (Std.Dev.)\tabularnewline
\hline
\hline 
1MB1&
2.1&
98&
27 (28)&
0.292&
0.352 (0.012)&
0.325 (0.0l0)\tabularnewline
\hline 
1BYW&
2.6&
110&
50 (45)&
0.292&
0.400 (0.046)&
0.321 (0.016)\tabularnewline
\hline 
1KXB&
2.9&
158&
54 (36)&
0.283&
0.351 (0.022)&
0.336 (0.006)\tabularnewline
\hline 
1RN7&
2.5&
112&
33 (29)&
0.270&
0.342 (0.032)&
0.328 (0.008)\tabularnewline
\hline 
2DBO&
2.76&
148&
57 (39)&
0.289&
0.370 (0.020)&
0.335 (0.017)\tabularnewline
\hline
\end{tabular}\end{center}

\label{frameloopTable}
\end{table}

\subsection{Variation of $R_{free}$ with collection size}

\label{collectionRfreeSection}

We define a collection as a set of independently refined structures
which when taken together may capture some aspects of structural heterogeneity.
This term is introduced to distinguish the collection from an ensemble
(see Section \ref{loopCollEnsSection}) which is also a set of structures,
but refined in an interdependent manner.

\begin{figure}

\caption{Variation of $R_{free}$ with collection size in whole chain, loop
and framework-loop exercises. For collection size of 0, baseline $R_{free}$
is shown. For collection size 1, the mean and standard deviation of
best $R_{free}$ models in CNS/{\small \rappertk} trajectories are
shown. The rest are calculated by combining the best $R_{free}$ individual
models with partial occupancies.}

\begin{center}\begin{tabular}{cc}
\includegraphics[%
  width=90mm,
  angle=90]{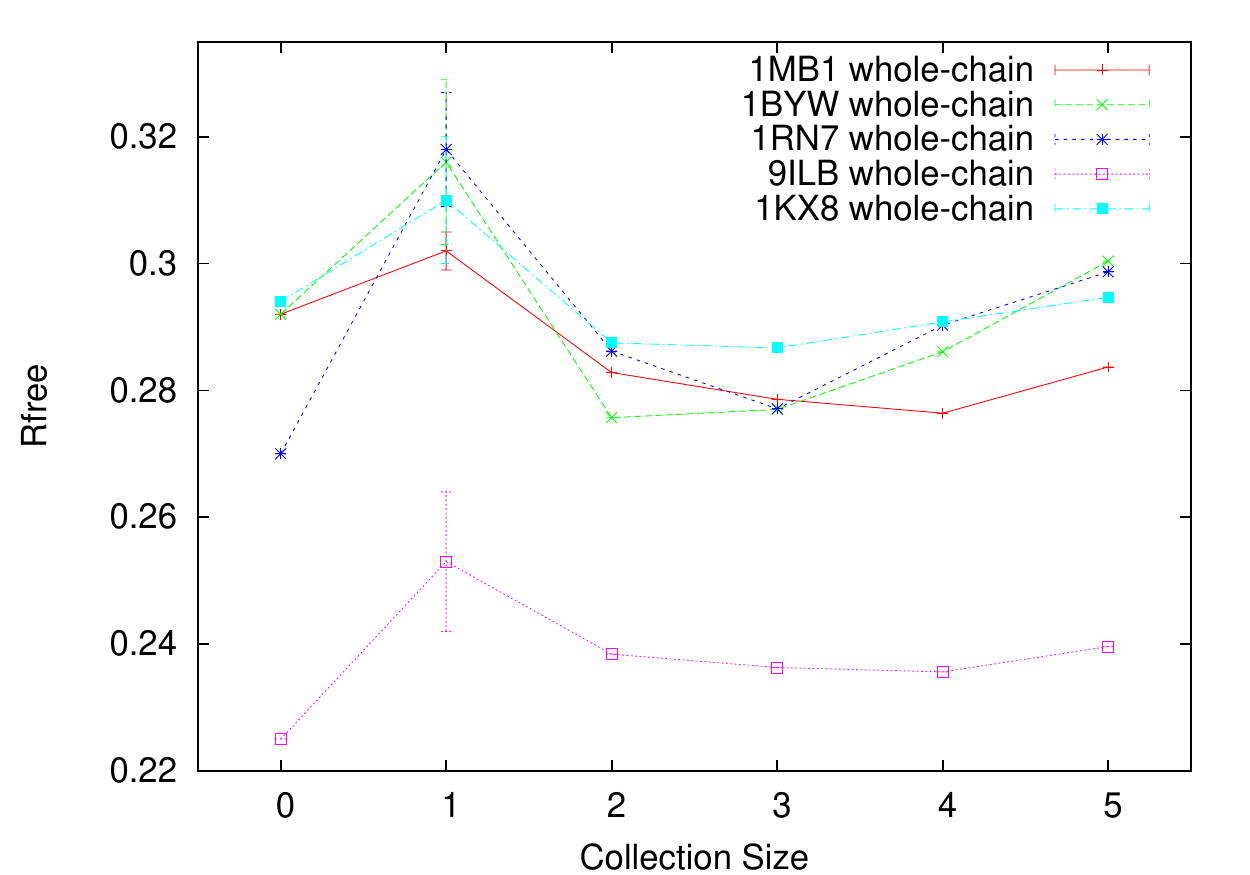}&
\includegraphics[%
  width=90mm,
  angle=90]{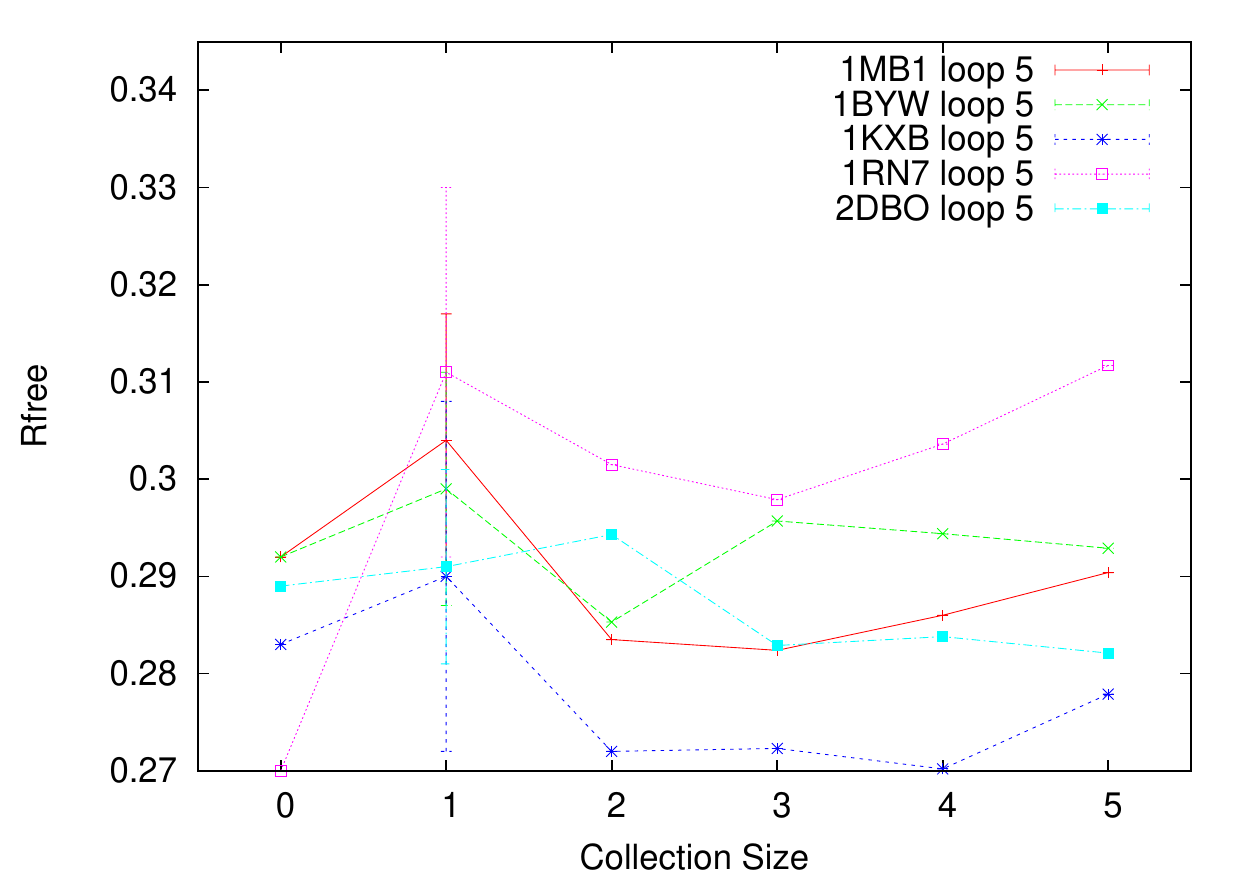}\tabularnewline
\includegraphics[%
  width=90mm,
  angle=90]{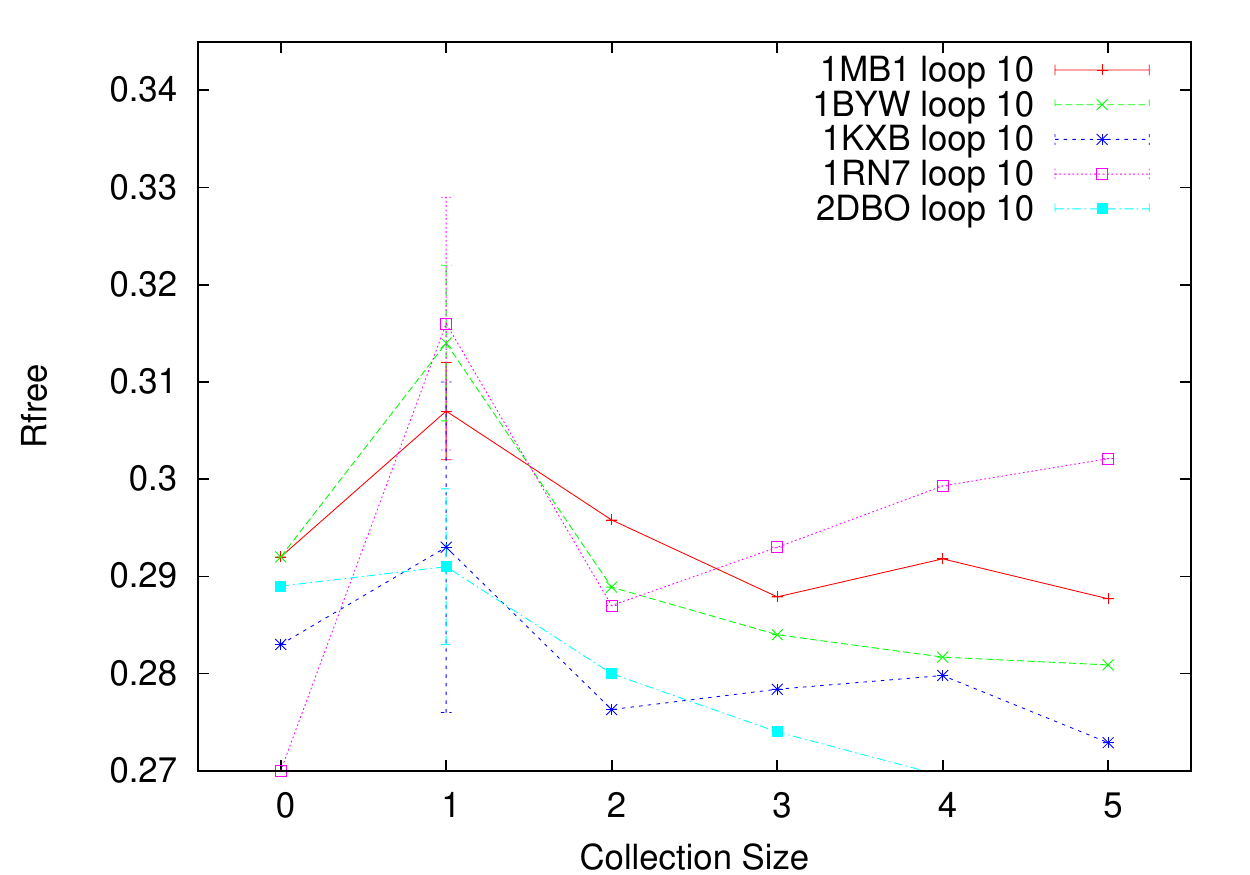}&
\includegraphics[%
  width=90mm,
  angle=90]{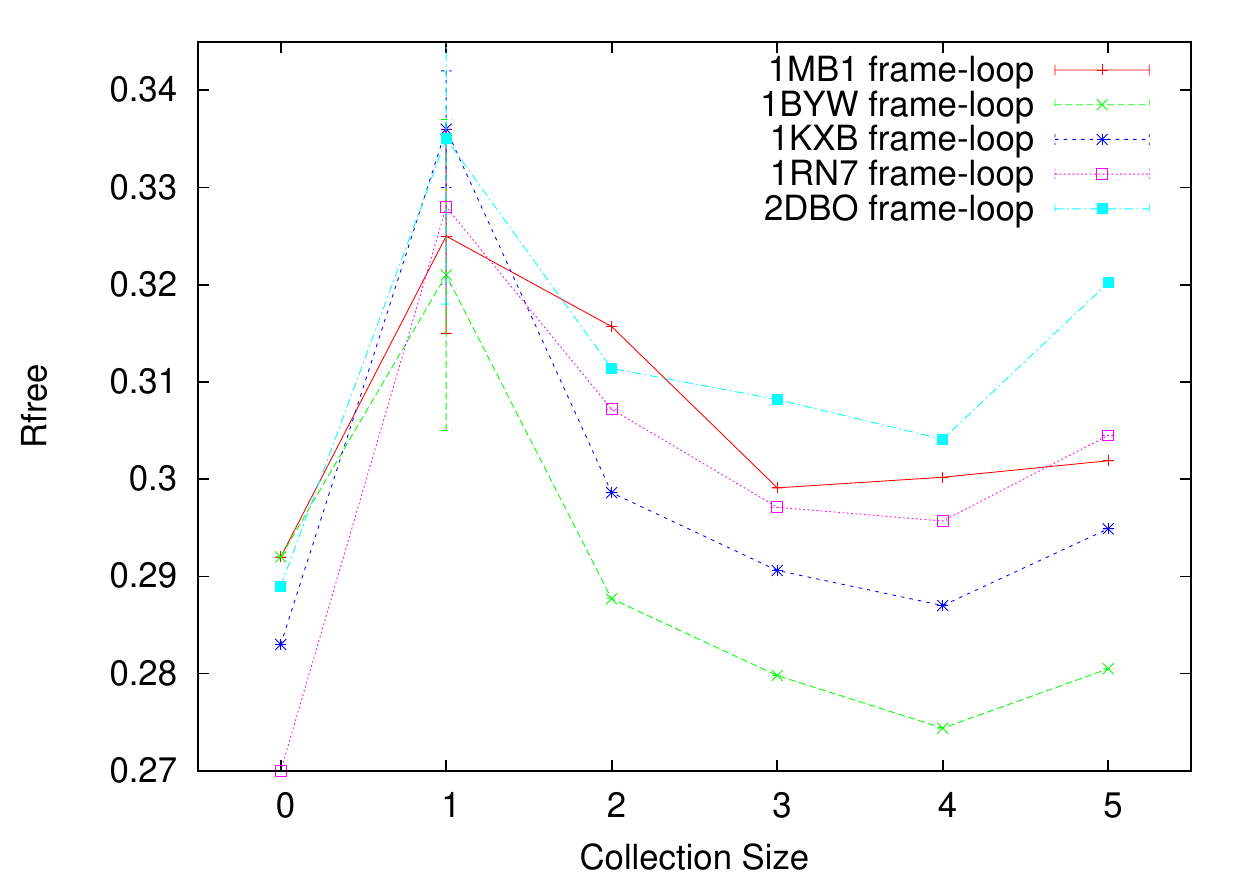}\tabularnewline
\end{tabular}\end{center}

\label{LoopCollectionRfree}
\end{figure}

For the previous exercises of refinement (whole chain, \Ang{5} loop,
\Ang{10} loop and loops with framework), single best $R_{free}$
models from five {\small \cns}/{\small \rappertk} trajectories are
chosen and combined to create collections, e.g. a three model collection
is created by choosing three lowest $R_{free}$ models from the five
and assigning occupancy of $0.33$ to each of them. A collection is
subjected to a short {\small \cns} refinement and $R_{free}$ at
its end is noted as collection $R_{free}$. Fig.\ref{LoopCollectionRfree}
shows the variation of such $R_{free}$ values as a function of collection
size. The drop in $R_{free}$ is modest and the highest when going
from collection size of 1 to 2 or 3. The collection $R_{free}$ generally
rises for sizes 4 and 5. This indicates the danger of overfitting
due to increase in number of parameters. Thus a straightforward combination
of models does not seem to be the correct way of describing heterogeneity.
Intelligent schemes for parameter reduction need to be investigated
in this regard, such as upper-bounding $B$-factors, enforcing the
same $B$-factor on corresponding atoms across all models and positionally
constraining them together when large variability is not expected.

\subsection{Mistakes in the composite protocol}

Although the {\small \cns}/{\small \rappertk} refinement produces
a well-refined structure very close to the baseline $R_{free}$, it
never betters the latter. This is due to imperfections in various
components of the protocol. Identification of residues to rebuild
relies on the $\chi^{2}$ correlation coefficient which can sometimes
be an unsatisfactory substitute for human judgement. This can lead
to unnecessary resampling of satisfactory bands and sometimes incorrect
conformers are not detected. The copying of non-protein atoms from
one round to next may sometimes result in their permanent misplacement.
The sampling problem may result in unsatisfactory bands because the
right conformation must be generated in order to be picked by the
electron density ranker. On the other hand, the ranker may not score
a correct conformation as the best one. If the density for a band
is weak, band sampling may sometimes lead into the density of waters
or ligands. In spite of these difficulties, the refinement statistics
presented in previous sections are satisfactory.

But when restraint radii are increased beyond those used here, the
serious problem of spatial overlap between restraint spheres of two
different bands starts affecting the band sampling. The correct density
for a band then may be occupied by a band sampled before it. In such
case, the correct density is always occupied by the wrong band. For
the wrong band, subsequent {\small \cns} refinement may take it so
far from its correct location that the restraint radii may be too
small to let the band be built correctly again. Further work is required
to get rid of the problem of band overlaps, either by restraint adjustments
or change in the sampling strategy.

\subsection{Modelling loop heterogeneity with collections and ensembles}

\label{loopCollEnsSection}

Conformational diversity is most pronounced in loop regions due to
the relatively smaller number of non-covalent interactions to maintain
order. Absence of good density for a loop when rest of the structure
has good density is a sure indication of the loop's flexibility. Modelling
heterogeneity is challenging because density is generally more confusing
for such regions owing to conformer overlaps and subsequent dilution
of shape information.

Heterogeneity can be modelled with collections or ensembles. Members
of a collection are single-conformer isotropic B-factor models determined
independently of one another. Members of an ensemble are determined
in a highly interdependent manner and have partial occupancies.

Derivation of a collection is a simple way to estimate the unavoidable
uncertainty in structure determination, but it cannot be said to represent
any structural correlations. A major advantage of collections is their
simplicity. A procedure that produces a single model can be executed
multiple times with different random seeds or starting models to generate
a collection. Thus the time taken increases only linearly as the collection
size.

An implicit assumption in the ensemble representation is that the
members are in dynamic equilibrium, making ensemble a much stronger
statement than the collection. Determining ensembles is very challenging
because it is unclear how to determine the number and occupancies
of the ensemble members prior to or during the refinement process.
Another major challenge is the linear increase in the number of parameters
which results in an exponential increase in search space and execution
time.

In order that its output be credible, any procedure that aims to model
the structural heterogeneity must be first validated using artificial
data where the \emph{real} heterogeneity is accurately known. To that
end, we have chosen a significantly simplified kind of heterogeneity
by generating artificial diffraction data in which the underlying
heterogeneity is restricted to a single loop and consists of two equally-occupied
loop conformations. For a loop each from 1MB1, 1BYW and 2DBO, two
conformers were generated by perturbing the loop to within \Ang{3}
$C_{\alpha}$ restraints and no sidechain restraints. All non-protein
atoms (ions, waters etc.) have been removed so as to reduce the density
dilution. Artificial diffraction data were created with the same cell,
spacegroup and resolution as the deposited structure. For self-consistency
in the {\small \cns} forcefield, data was generated iteratively.
The average of the two conformations was considered as the starting
conformation for further heterogeneity modelling.

A collection of 4 members was generated for each loop with {\small \cns}/{\small \rappertk}
protocol used previously. An ensemble consisting of 2 members was
generated for each loop using multiconformer sampling described previously.
As with the single-conformer protocol, multiconformers were sampled
iteratively and alternatingly with {\small \cns}. Enrichment was
increased to 10 and population size to 200 to be able to build reasonable
models. Positive electron density restraint was enforced on mainchain.

\begin{figure}

\caption{Heterogeneity modelling with collection and ensemble for a loop in
PDB 1MB1. The top panel shows the artificially generated loop heterogeneity
with corresponding electron density contoured at $1\sigma$. The middle
and bottom panels respectively show a 4-member collection and 2-member
ensemble model of that heterogeneity.}

\begin{center}\includegraphics[%
  width=60mm,
  angle=90]{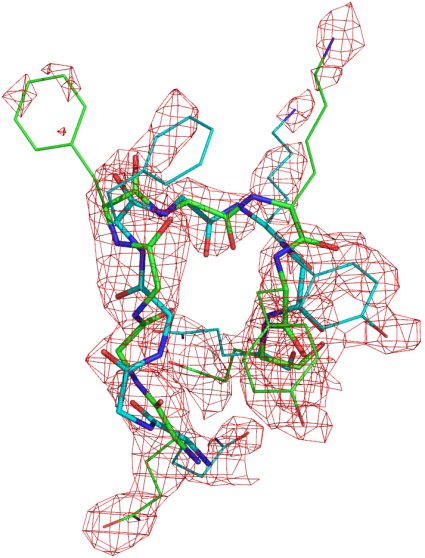}\end{center}

\begin{center}\includegraphics[%
  width=60mm,
  angle=90]{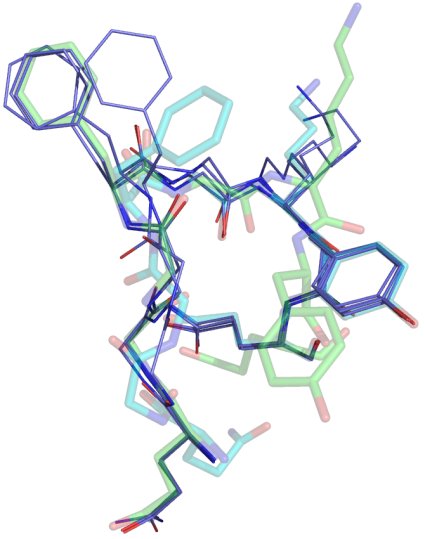}\end{center}

\begin{center}\includegraphics[%
  width=60mm,
  angle=90]{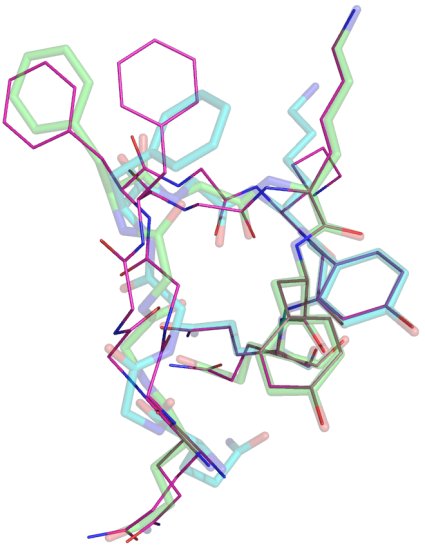}\end{center}

\label{EC1mb1}
\end{figure}

\begin{figure}

\caption{Heterogeneity modelling with collection and ensemble for a loop in
PDB 1BYW. Panels arranged as in Fig.\ref{EC1mb1}.}

\begin{center}\includegraphics[%
  width=60mm,
  angle=90]{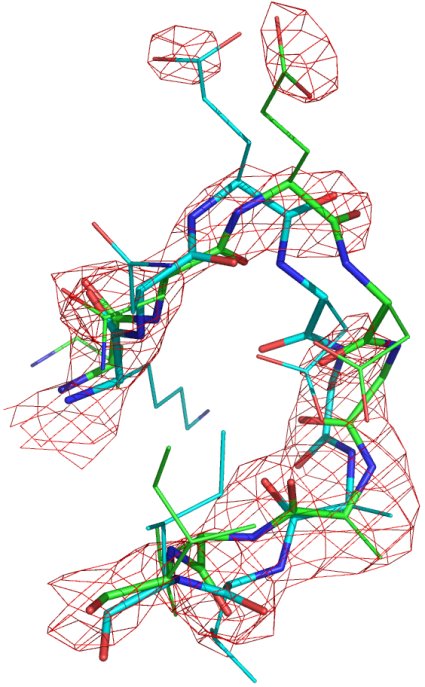}\end{center}

\begin{center}\includegraphics[%
  width=60mm,
  angle=90]{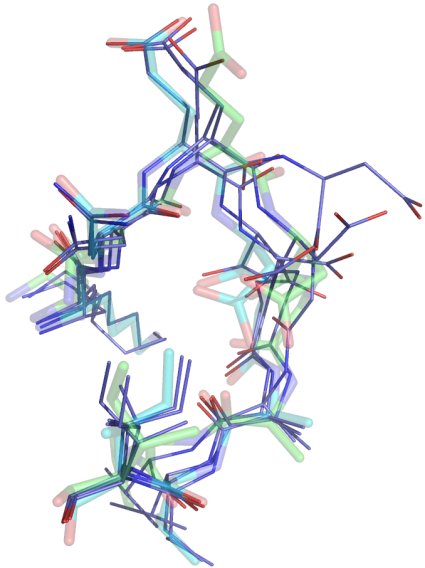}\end{center}

\begin{center}\includegraphics[%
  width=60mm,
  angle=90]{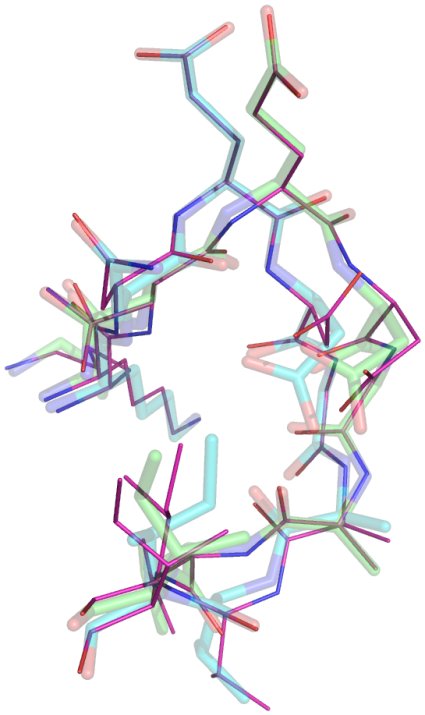}\end{center}

\label{EC1byw}
\end{figure}

\begin{figure}

\caption{Heterogeneity modelling with collection and ensemble for a loop in
PDB 2DBO. Panels arranged as in Fig.\ref{EC1mb1}. In the top panel,
the red contours correspond to $0.5\sigma$.}

\begin{center}\includegraphics[%
  width=60mm,
  angle=90]{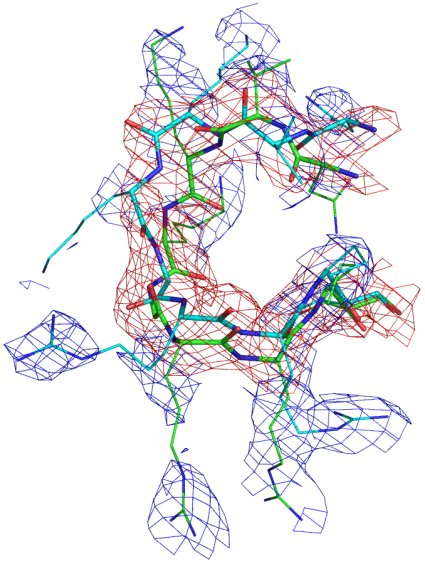}\end{center}

\begin{center}\includegraphics[%
  width=60mm,
  angle=90]{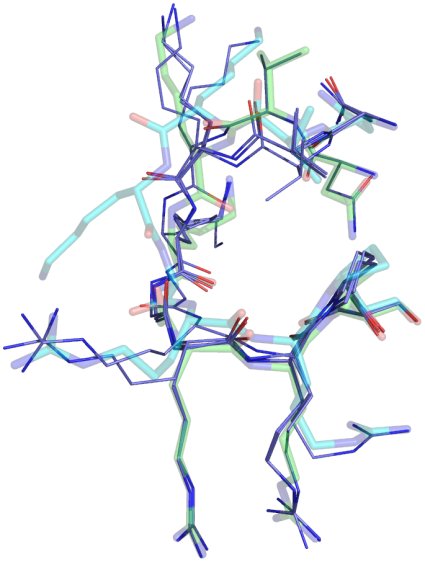}\end{center}

\begin{center}\includegraphics[%
  width=60mm,
  angle=90]{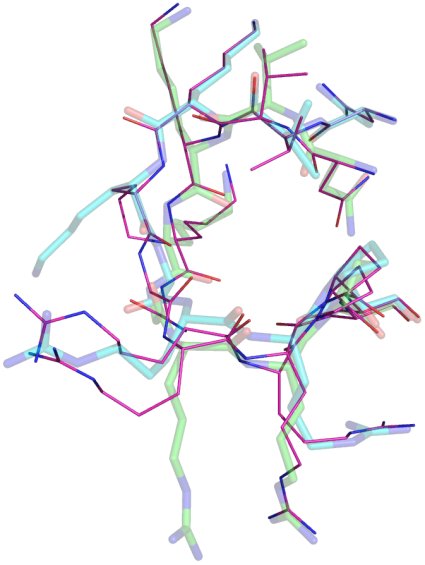}\end{center}

\label{EC2dbo}
\end{figure}

The performance of collections and ensembles can be visually inspected
in Fig.\ref{EC1mb1}, Fig.\ref{EC1byw} and Fig.\ref{EC2dbo} but
it is essential to quantify quality of heterogeneity modelling objectively.
The two important quantities to measure are: the extent to which both
conformers are captured and the extent to which at least one conformer
is captured. The former (multiconformer quality index, MQI) should
quantify how much of the underlying diversity is represented and the
latter (single-conformer quality index, SQI) should quantify how well
at least one of the heterogenous states is modelled.

If conformers $H_{i}$ constitute the true underlying heterogeneity
and $M_{i}$ are the ensemble or collection members which model it,
then MQI and SQI can be calculated as:

\begin{eqnarray}
MQI & = & \sum_{i}min_{j}(Rmsd(H_{i},M_{j}))\label{eqMqiSqi}\\
SQI & = & min_{i,j}(Rmsd(H_{i},M_{j}))\nonumber \end{eqnarray}

where {\small \rmsd} is calculated over the atoms of interest, e.g.
a sidechain or all $C_{\alpha}$ atoms in the loop. Note that these
expressions do not consider the occupancies. Table \ref{ECtable}
quantifies the performance of ensembles and collections using $R_{free}$,
MQI over loop $C_{\alpha}$ atoms, and MQI, SQI over each sidechain.
The ensemble $R_{free}$ values are smaller than those for collections
as expected due to greater number of parameters. $C_{\alpha}$ MQI
suggests that mainchain heterogeneity is modelled better in the ensemble
method. Sidechain SQIs do not show any systematic difference between
the two methods, which suggests that both methods capture the rotameric
heterogeneity to a similar extent. But sidechain MQIs tend to be slightly
better for ensemble than collection. This is not surprising because
in principle, the only limitations on the ensemble method are sampling
and ranking of conformational extensions. Generally the higher density
option is chosen in single-conformer modelling but a lower density
can also be chosen in multiconformer modelling due to a greater number
of atoms to place in the density. This is evident from residues Lys-72
and Tyr-73 in 1MB1, Glu-118, Asp-119 in 1BYW and Lys-86, Lys-87 in
2DBO where collection models are biased towards one of the loop conformations
due to weak density.

\begin{table}

\caption{Comparison of collection and ensemble heterogeneity modelling styles
with artificially generated 2-conformer heterogeneity for 3 loops.
MQI and SQI are in \Ang{} units.}

\begin{center}\begin{sideways}
\begin{tabular}{|c|c|c|c|c|c|c|}
\hline 
PDB&
Baseline&
Collection $R_{free}$&
Ensemble&
MQI-$C_{\alpha}$&
MQI-$C_{\alpha}$&
Residue-wise MQI, SQI for collection;\tabularnewline
&
$R_{free}$&
Mean (Std.Dev.)&
$R_{free}$&
Collection&
Ensemble&
MQI, SQI for ensemble\tabularnewline
\hline
\hline 
1MB1&
0.055&
0.127 (0.003)&
0.094&
2.11&
1.42&
\begin{tabular}{ccccc}
&
&
&
&
\tabularnewline
\hline
Gln-67&
4.269&
0.354&
4.285&
0.516\tabularnewline
\hline
Phe-70&
3.575&
0.262&
3.271&
0.670\tabularnewline
\hline
Lys-72&
5.624&
0.889&
2.539&
0.235\tabularnewline
\hline
Tyr-73&
3.316&
0.151&
0.233&
0.115\tabularnewline
\hline
Gln-74&
1.606&
0.349&
1.614&
0.189\tabularnewline
\hline
&
&
&
&
\tabularnewline
\end{tabular}\tabularnewline
\hline 
1BYW&
0.159&
0.242 (0.004)&
0.192&
0.79&
0.58&
\begin{tabular}{ccccc}
&
&
&
&
\tabularnewline
\hline 
Lys-116&
0.994&
0.325&
1.004&
0.275\tabularnewline
\hline 
Asn-117&
1.849&
0.529&
1.861&
0.275\tabularnewline
\hline 
Glu-118&
2.171&
0.169&
0.178&
0.049\tabularnewline
\hline 
Asp-119&
4.580&
1.833&
3.703&
1.281\tabularnewline
\hline 
Val-122&
1.402&
0.279&
1.000&
0.262\tabularnewline
\hline 
Ile-123&
0.829&
0.273&
2.559&
0.677\tabularnewline
\hline 
&
&
&
&
\tabularnewline
\end{tabular}\tabularnewline
\hline 
2DBO&
0.292&
0.312 (0.005)&
0.271&
2.01&
1.10&
\begin{tabular}{ccccc}
&
&
&
&
\tabularnewline
\hline 
Asn-84&
0.696&
0.286&
1.710&
0.147\tabularnewline
\hline 
Val-85&
0.598&
0.210&
3.202&
1.073\tabularnewline
\hline 
Lys-86&
2.160&
0.901&
0.622&
0.153\tabularnewline
\hline 
Lys-87&
6.821&
0.456&
3.830&
0.451\tabularnewline
\hline 
Arg-89&
0.814&
0.136&
7.495&
1.480\tabularnewline
\hline
Arg-90&
1.739&
0.394&
2.174&
0.442\tabularnewline
\hline
Pro-91&
0.850&
0.220&
1.273&
0.522\tabularnewline
\hline
Ser-92&
0.396&
0.173&
0.576&
0.189\tabularnewline
\hline
&
&
&
&
\tabularnewline
\end{tabular}\tabularnewline
\hline
\end{tabular}
\end{sideways}\end{center}

\label{ECtable}
\end{table}

\section{Concluding Discussion}

The problem of automated crystallographic refinement is interesting,
challenging and has significant immediate practical relevance. An
automated solution for building single conformer models from approximate
spatial restraints, combined with an automated method to explore heterogeneity,
will allow the crystallographic community to revisit and annotate
the entire {\small \pdb} with structural variability information.
Such information will have an impact on all analyses that rely on
accurate coordinate information, like in-silico ligand design and
binding, non-covalent interactions and sequence-structure conservation,
etc. It will also significantly change the understanding of crystalline
state and protein flexibility, benefitting the refinement process.
The main components for successful heterogeneity annotation are reliable
construction of single-conformer models and reliable estimation of
heterogeneity that is predominantly seen in loops. This work has attempted
to develop methods to that end.

Application of the {\small \rapper} approach to crystallographic
refinement (\cite{rapperKnowledgeXray}, \cite{rapperLowResolution})
has the primary benefit of crossing the energy barriers in a non-random
manner, based on knowledge-based sampling instead of kinetic sampling.
As described in \cite{rappertk}, {\small \rapper} has been reformulated
as {\small \rappertk} recently, creating possibilities of applying
knowledge-based sampling in many different ways. In this work, we
showed that {\small \rappertk} can be used to automatically refine
the whole protein structure starting from reliable positional restraints
on mainchain and sidechain. This benchmarked its performance vis-a-vis
{\small \rapper} as reported by \cite{rapperKnowledgeXray} for a
similar task. Then we showed that by efficient use of available restraints,
single loops can be modelled in protein structures to native-like
quality with few positional restraints. The efficient use of available
restraints was a result of symmetry-related clashchecking, restraint
propagation using loop anchors and sampling from both anchors simultaneously.
The same strategy could be extended to a more realistic problem of
a large uncertainty in loop regions and an imperfect secondary structure
framework. The {\small \cns}-only refinements, run as controls, showed
the value added by {\small \rappertk} to the refinement task.

In addition to determination of single-conformer models under differing
restraint qualities, we started addressing the challenge of heterogeneity
assessment in loops. This is indeed a very difficult problem with
fundamental unknowns like number of conformers, correlations within
heterogeneity and relative occupancies. Conformational heterogeneity
can be divided into two types: the simpler sidechain-only heterogeneity
where mainchain is nearly the same and the all-atom heterogeneity
where mainchain also takes distinct conformations. The latter can
be further divided based on the extent of spatial overlap between
the conformers. Sidechain-only heterogeneity is relatively easy than
the all-atom heterogeneity because the density is likely to contain
good cues about diversity. But for overlapping conformers, a visual
inspection of density is less likely to be helpful. There are two
distinct ways to model heterogeneity, which we have termed collections
and ensembles, depending upon interdependence of member conformations.
For single-loop 2-conformer overlapping heterogeneity, we generated
both the collections and ensembles and assessed how well they modelled
the heterogeneity. The main observation was that the collection was
generally biased towards the higher electron density. The ensemble
method, due to more freedom and parameters available to it, manages
to avoid this trap and fit two distinct conformers, leading to better
modelling of heterogeneity than the collection.

Various pitfalls of the composite refinement protocol were recognized
and they need to be addressed in future. Addressing the problem of
overlapping bands will significantly increase the reliability of the
method given very approximate positional restraints, and make the
method more useful in low resolution, large uncertainty cases. Perhaps
many models can be generated for such bands and the best combination
of those models can be used. At lower resolution, use of coarse-grained
sampling (fragment sampling) may also be useful, followed by fine-grained
$\phi-\psi-\chi$ sampling.

From the heterogeneity perspective, a fundamental question to address
would be the estimation of the nature of underlying conformations
before attempting to model it because ensemble sampling must have
prior knowledge of the number and occupancies of its members. Generation
of collections seems the only way for such estimation, for which collection
modelling method will have to be modified suitably to sample within
electron density yet avoid the bias towards higher density. The main
challenge of ensemble sampling is the explosion in conformational
freedom and the work ahead will have to focus on efficiently scaling
this sampling method for larger ensemble size.

\bibliographystyle{marko}
\addcontentsline{toc}{section}{\refname}\bibliography{arxivLoops}

\end{document}